\documentclass[prb,preprint,amsmath,amssymb,superscriptaddress,notitlepage]{revtex4-1}
\usepackage{epsfig}
\usepackage{graphicx}
\usepackage{color}
\usepackage{bm}
\usepackage{upgreek}
\usepackage{soul}

\begin{document}

\renewcommand{\figurename}{Fig.}
\renewcommand{\tablename}{Tab.}


\title{Altermagnetism: spin-momentum locked phase protected by non-relativistic symmetries 
}
\author{Libor \v{S}mejkal}
\affiliation{Institut f\"ur Physik, Johannes Gutenberg Universit\"at Mainz, 55128 Mainz, Germany}
\affiliation{Institute of Physics, Czech Academy of Sciences, Cukrovarnick\'a 10, 162 00, Praha 6, Czech Republic}
\author{Jairo~Sinova}
\affiliation{Institut f\"ur Physik, Johannes Gutenberg Universit\"at Mainz, 55128 Mainz, Germany}
\affiliation{Institute of Physics, Czech Academy of Sciences, Cukrovarnick\'a 10, 162 00, Praha 6, Czech Republic}
\author{Tomas~Jungwirth}
\affiliation{Institute of Physics, Czech Academy of Sciences, Cukrovarnick\'a 10, 162 00, Praha 6, Czech Republic}
\affiliation{School of Physics and Astronomy, University of Nottingham, NG7 2RD, Nottingham, United Kingdom}

\maketitle

\subsection*{Abstract}

The search for novel magnetic quantum phases, phenomena and functional materials has been guided by relativistic magnetic-symmetry groups in coupled spin and real space\cite{Landau1984,Shubnikov1964,Tavger1956,Bradley,Litvin2013,Gallego2016a} from the dawn of the field  in 1950s\cite{Landau1984,Shubnikov1964,Tavger1956} to the modern era of topological matter\cite{Smejkal2017b,Watanabe2018c,Xu2020}. However, the magnetic groups  cannot  disentangle\cite{Brinkman1966,Andreev1980}  non-relativistic phases and effects, such as the recently reported unconventional spin physics in collinear antiferromagnets\cite{Smejkal2021b,Smejkal2020,Ahn2019,Hayami2019,Naka2019,Yuan2020,Feng2020a,Hayami2020,Reichlova2020,Yuan2021a,Egorov2021,Mazin2021,Gonzalez-Hernandez2021,Naka2021,Bose2021,Bai2021,Smejkal2021,Shao2021}, from the typically weak relativistic spin-orbit coupling phenomena. Here we discover that more general spin symmetries in decoupled spin and crystal space\cite{Brinkman1966,Litvin1974,Litvin1977} categorize non-relativistic collinear magnetism in  three phases: conventional ferromagnets and antiferromangets, and a third distinct phase combining zero net magnetization with an alternating spin-momentum locking in energy bands, which we dub "altermagnetic". For this third basic magnetic phase, which is omitted by the relativistic magnetic groups, we develop a spin-group theory  describing  six characteristic types of the altermagnetic spin-momentum locking. We demonstrate an extraordinary spin-splitting mechanism in altermagnetic bands originating from a local electric crystal field, which contrasts with the conventional magnetic or relativistic splitting by global magnetization or inversion asymmetry.
Based on first-principles calculations, we  identify altermagnetic candidates ranging from insulators and metals to a parent crystal of cuprate superconductor.  Our results  underpin   emerging  research of quantum phases and spintronics in high-temperature magnets with  light elements, vanishing net magnetization,  and strong spin-coherence.


\vspace*{6.8cm}

\pagebreak

\subsection*{Main}

The conventional symmetry theory of magnetic crystals considers transformations in coupled real physical space and the space of magnetic moment vectors. In other words, the rotation transformations acting on the coordinates of the atoms, subject to the standard crystallographic restrictions, simultaneously act on the components of the magnetic moment vectors\cite{Litvin1974,Litvin1977,Bradley}. This symmetry formalism naturally arises  from the classical orbital-current model of magnetic moments  \cite{Landau1984}, as well as from the relativistic quantum-mechanical description of coupled spin and orbital degrees of freedom of electrons \cite{Landau1984,Turov1965}. The corresponding magnetic groups \cite{Bradley} have been broadly applied in the research of equilibrium and non-equilibrium phenomena, including their modern topological variants \cite{Smejkal2017b,Watanabe2018c,Xu2020}, and have represented the primary tool for a systematic classification of hundreds of magnetic structures in materials' databases \cite{Litvin2013,Gallego2016a}.

A spin-group formalism is a vast generalization of the conventional magnetic groups \cite{Brinkman1966,Litvin1974,Litvin1977}. It considers pairs of transformations $[R_1\parallel R_2]$, where the transformations on the left of the double vertical bar act only on the spin space and on the right of the double vertical bar  only on the real space \cite{Brinkman1966,Litvin1974,Litvin1977}. The symmetry landscape of the spin groups is much richer because, in general, different rotation transformations can simultaneously act on the spin and real space, and only the transformations in the real space are crystallographically-restricted. Despite their richness, studies based on the spin symmetries have appeared only sporadically in literature. For example, they were applied for the classification of possible spin arrangements on crystals, with an emphasis on complex non-collinear  or disordered structures \cite{Andreev1980}. Overall, however, the spin-group formalism has  remained unexploited.

The key significance of  the spin groups is that they can offer a systematic symmetry description of  physics arising from non-relativistic electro-magnetic crystal potentials, which typically play the leading role in magnetism \cite{Brinkman1966,Andreev1980}. Here by the electric crystal potential we refer to the internal potential in the non-magnetic phase of the crystal, as described, e.g., by the local-density approximation of the density functional theory (DFT); by the additional magnetic component we refer to the modification of the internal crystal potential due to the transition to the magnetically ordered phase. Since the magnetic groups represent only a small subset of the spin groups \cite{Litvin1977}, they are prone to omit prominent magnetic phases dominated by the non-relativistic electro-magnetic crystal potentials. 
For example, the magnetic groups can only determine whether a net magnetization is allowed or not, but do not distinguish  ferromagnets from antiferromagnets in which magnetization arises only as a weak relativistic perturbation\cite{Landau1984}. For the band structures, the magnetic groups were only used to identify a violation of Kramers spin-degeneracy, again without disentangling the non-relativistic and relativistic origin \cite{Yuan2021,Egorov2021a}. 

The central topic of our work is the altermagnetic phase of collinear magnets, which we delimit using the non-relativistic spin groups from a ferromagnetic phase with non-zero net magnetization, and from a Kramers spin-degenerate antiferromagnetic phase. Altermagnets have split, but equally populated spin-up and spin-down energy iso-surfaces. The corresponding planar or bulk spin-momentum locking is protected by the spin-group symmetries, and can be characterized by an even-integer spin winding number.  

We discover an extraordinary microscopic spin-splitting mechanism in altermagnetic materials which originates from a local anisotropic electric crystal field, i.e., from crystal properties of the non-magnetic phase. It is fundamentally distinct from the earlier considered internal magnetic-interaction mechanisms \cite{Pekar1965,Yuan2020,Ahn2019,Hayami2020,Smejkal2021}, such as the anisotropic spin-dependent hopping in the magnetic state\cite{Hayami2020,Smejkal2021}. The altermagnetic spin-splitting by the local electric crystal field also starkly contrasts with the conventional mechanisms of the ferromagnetic splitting due to the global magnetization, or the spin-orbit splitting due to the global inversion asymmetry.   It opens a  third paradigm for designing spin quantum phases of matter based on the strong crystal-field effects, complementing the widely explored relativistic or many-body phenomena\cite{Sobota2021}.

The spin-groups describing the altermagnetic phase have no correspondence in the magnetic groups. The recent reports\cite{Smejkal2021b,Smejkal2020,Ahn2019,Hayami2019,Naka2019,Yuan2020,Feng2020a,Hayami2020,Reichlova2020,Yuan2021a,Egorov2021,Mazin2021,Gonzalez-Hernandez2021,Naka2021,Bose2021,Bai2021,Smejkal2021,Shao2021}  pointing towards  uncharted Berry phase or charge-spin conversion physics 
then not only highlight the range of potential science and technology implications of this new magnetic phase, but also the importance of its proper symmetry description by the non-relativistic spin groups. 

\subsection*{Derivation of spin group categorization of non-relativistic collinear magnetism}

We start with the derivation of three distinct spin group types by which we catagorize all non-relativistic collinear magnets into the ferromagnetic, Kramers spin-degenerate antiferromagnetic, and altermagnetic  phases. In general, spin groups can be expressed as a direct product of so-called spin-only groups, containing  transformations of the spin space alone, and so-called  non-trivial spin groups containing the elements $[R_1\parallel R_2]$, but no elements of the spin-only group \cite{Litvin1974,Litvin1977}. For the collinear spin arrangements on crystals, the spin-only group is given by\cite{Litvin1974,Litvin1977}  ${\bf C}_\infty + \bar{C}_2{\bf C}_\infty$.  Here ${\bf C}_\infty$  is a group representing all rotations of the spin space around the common axis of spins, and $\bar{C}_2$ is  a 180$^\circ$ rotation around an axis perpendicular to the spins, combined with the spin-space inversion.  We emphasize that because of the absence of the spin-only group symmetries in magnetic groups, the collinear spin groups have no correspondence in the magnetic groups. In other words, unlike the spin groups, the conventional relativistic magnetic groups do not strictly delimit collinear magnets.

Next we point out that the spin-space inversion in the spin groups enters via the time-inversion \cite{Litvin1974,Litvin1977,Andreev1980}. Since time is common in both spin and real space, the spin-space inversion is accompanied by a simultaneous time-inversion  in the real space (${\cal T}$). Although ${\cal T}$ acts as  identity on the real space coordinates of the atoms, it flips the sign of the crystal momentum. This is important for the band-structure spin symmetries. In particular, we will now use the symmetry $[\bar{C}_2\parallel {\cal T}]$,  which follows directly from the above spin-only group of the collinear magnets, and from the simultaneous action of the time-inversion on the spin and real (momentum) space. When applying the transformation $[\bar{C}_2\parallel {\cal T}]$ on spin ($s$) and crystal-momentum ({\bf k}) dependent bands $\epsilon(s,{\bf k})$, we obtain $[\bar{C}_2\parallel {\cal T}]\epsilon(s,{\bf k})=\epsilon(s,-{\bf k})$. Next, since $[\bar{C}_2\parallel {\cal T}]$ is a symmetry of non-relativistic collinear spin arrangements on crystals, $[\bar{C}_2\parallel {\cal T}]\epsilon(s,{\bf k})=\epsilon(s,{\bf k})$,  and hence $\epsilon(s,{\bf k})=\epsilon(s,-{\bf k})$. The bands are invariant under real-space (crystal-momentum) inversion even if the crystal lacks the real-space inversion symmetry.

We now move on to the  non-trivial spin groups  which are obtained by combining the groups of real-space crystallographic transformations with the groups of spin-space transformations  \cite{Litvin1974,Litvin1977}. Regarding the real-space crystallographic groups, we can limit our basic classification to the Laue groups,  which are the crystallographic point groups containing the real-space inversion symmetry element. The corresponding spin Laue groups describe the non-relativistic spin-momentum locking, which is present in the band structure independent of the crystal's real-space translation symmetries, and independent of whether the crystal does or does not have the real-space inversion symmetry. The independence of translations is straight forward, and  the invariance of  bands under real-space (crystal-momentum) inversion was derived above from the spin-only group symmetry of the collinear spin arrangements.  (Note that additional symmetry  analysis of  complex band structures, based on the direct extension of the theory to spin space groups, is beyond the scope of the present work.)

Regarding the groups of spin-space transformations, there can be some freedom in their choice\cite{Litvin1974,Litvin1977}. For the collinear spin arrangements, one of the two spin-space transformation groups is ${\bf S}_1=\{E\}$, i.e., contains just the spin-space identity\cite{Litvin1974}. We choose the second group in the form of ${\bf S}_2=\{E,C_2\}$ which is favorable for our derivation of the categorization into the three phases of non-relativistic collinear magnets. The group contains the spin-space identity and the 180$^\circ$ rotation of the spin space around an axis perpendicular to the spins. 
(We note that because of the above spin-only group symmetry element $\bar{C}_2$, and because the product of spin-space transformations $\bar{C}_2{C}_2$ is equal to the spin-space inversion, an  alternative choice\cite{Litvin1974} of ${\bf S}_2$ contains the spin-space inversion instead of $C_2$.) 

We construct all the non-trivial spin (Laue) groups, whose elements on the left of the double vertical bar form a group of the spin-space transformations and on the right of the double vertical bar a (Laue) group of the real-space crystallographic transformations, by using the isomorphism theorem\cite{Litvin1974}. It  implies the procedure of combining all isomorphic coset decompositions of the two groups, i.e., decompositions  with the same number of cosets for the two groups\cite{Litvin1974}. (A coset decomposition of a group {\bf X} is given by ${\bf X}={\bf x} + X_1{\bf x}+ X_2{\bf x} + \dots$, where {\bf x} is a subgroup of {\bf X} and $X_i$ are elements of {\bf X} \cite{Litvin1974}.) The details of our derivation are in Supplementary Sec.~I. Here we summarize the result in which  all the non-trivial spin Laue groups describing $\epsilon(s,{\bf k})$ of collinear magnets are arranged into the following three distinct types using the isomorphic coset decompositions. 

The first type of the non-trivial spin Laue groups  is given by ${\bf R}_s^{\rm I}=[E\parallel{\bf G}]$, where ${\bf G}$ is the crystallographic Laue group. Because there are 11 different crystallographic Laue groups, there are also 11 different ${\bf R}_s^{\rm I}$ groups. As highlighted in Fig.~1, the ${\bf R}_s^{\rm I}$ groups do not imply spin degeneracy of  $\epsilon(s,{\bf k})$ at any {\bf k}-point. They describe non-relativistic band structures of collinear ferromagnets. 

The  second type of the non-trivial spin Laue groups is given by ${\bf R}_s^{\rm II}=[E\parallel{\bf G}]+[{C}_2\parallel{\bf G}]$. Here the $[{C}_2\parallel E]$ symmetry (recall that ${\bf G}$ is a group containing the real-space identity $E$ element) implies spin-degeneracy of $\epsilon(s,{\bf k})$ for all {\bf k}-vectors in the Brillouin zone.  The 11 different ${\bf R}_s^{\rm II}$ groups describe non-relativistic band structures of non-magnetic crystals, as well as Kramers spin-degenerate bands of collinear antiferromagnets (see Fig.~1). The corresponding antiferromagnetic spin arrangements on crystals have  a symmetry $[{C}_2\parallel {\bf t}]$, which interchanges atoms and rotate the spin by 180$^\circ$ between opposite-spin sublattices. Here ${\bf t}$ on the right side of the double vertical bar is a real-space translation.  Examples\cite{Marti2014,Li2019a} are antiferromagnets FeRh or MnBi$_2$Te$_4$.  The ${\bf R}_s^{\rm II}$ groups also describe Kramers spin-degenerate collinear antiferromagnetism in crystals with the opposite-spin-sublattice transformation symmetry  $[{C}_2\parallel \bar{E}]$, where $\bar{E}$ on the right side of the double vertical bar is the real-space inversion (see Supplementary Sec.~I). Here  examples\cite{Smejkal2016,Elmers2020} are antiferromagnets CuMnAs or Mn$_2$Au.


The remaining third distinct type of the non-trivial spin Laue groups is given by 
\begin{equation}
{\bf R}_s^{\rm III}=[E\parallel{\bf H}]+[{C}_2\parallel A] \, [E\parallel{\bf H}]=[E\parallel{\bf H}]+[{C}_2\parallel{\bf G-H}],
\label{Rsstar}
\end{equation}
with $A$ representing a real-space proper or improper rotation  which interchanges atoms between opposite-spin sublattices.
We see from Eq.~(\ref{Rsstar}) that for ${\bf R}_s^{\rm III}$, {\bf G} is expressed as a sublattice coset decomposition, where  {\bf H} contains only the real-space transformations which interchange atoms between same-spin sublattices, and ${\bf G}-{\bf H}$ contains  only the real-space transformations which interchange atoms between opposite-spin sublattices.    Lifted spin-degeneracies in  the ${\bf R}_s^{\rm III}$   groups are allowed for crystal momenta 
whose little group does not contain ${\bf H}A$ elements. They satisfy 
${\bf H}A \, {\bf k}={\bf k}^\prime\neq{\bf k}$, implying that $\epsilon(s,{\bf k})=[C_2\parallel {\bf H}A]\epsilon(s,{\bf k})=\epsilon(-s,{\bf k}^\prime)$ (see Fig.~1). It guarantees that the spin-up and spin-down energy iso-surfaces are split, but have the same number of states. These non-relativistic band structure signatures of the ${\bf R}_s^{\rm III}$ altermagnetic phase are unparalleled in the ${\bf R}_s^{\rm I}$ ferromagnetic or ${\bf R}_s^{\rm II}$ Kramers-degenerate antiferromagnetic phases. Simultaneously, there are 10 different ${\bf R}_s^{\rm III}$ groups which is comparable to the number of the ${\bf R}_s^{\rm I}$ or ${\bf R}_s^{\rm II}$ groups, suggesting that altermagnetism is abundant. The 10 non-trivial spin Laue groups of altermagnets are listed in Tab.~1, where we adopted Litvin's notation of the spin groups \cite{Litvin1977}. Note, that they are constructed from only 8 different crystallographic Laue groups. On the hand, the altermagnetic spin Laue groups cannot be constructed for the 3 remaining crystallographic Laue groups, namely from ${\bf G}=\bar{1}$, $\bar{3}$ or $m3$.

Before moving to the analysis of the altermagnetic spin-momentum locking protected by the symmetries of the spin-group, we empasize additional differences from the magnetic groups. The latter are constructed by combining crystallographic groups (with the same transformations acting simultaneously on coordinates of atoms and components of magnetic moment vectors) with one group containing the identity element alone, and a second group containing the identity and the time-inversion. Comparing this to the spin-group formalism with ${\bf S}_1$ also containing only the identity  element and ${\bf S}_2$ with again two elements implies, that for describing all magnetic structures, the relativistic symmetry formalism has the same number of different magnetic groups as is the number of different spin groups describing exclusively collinear spin arrangements. Also, because of the crystallographic operations applied in the coupled real and spin space, there is no counterpart in magnetic groups of the sublattice coset decomposition form of the altermagnetic ${\bf R}_s^{\rm III}$ spin groups (see Supplementary Sec.~I). As we further highlight below, the decomposition into same-spin and opposite-spin-sublattice transformations  in ${\bf R}_s^{\rm III}$ plays a central role in the understanding of altermagnetism. 

\subsection*{Altermagnetic spin-momentum locking protected by spin symmetries}
We now discuss basic characteristics of the spin-momentum locking in altermagnets.  
The $\boldsymbol\Gamma$-point is invariant under all real-space transformations. The $[{C}_2\parallel A]$ symmetry present in the ${\bf R}_s^{\rm III}$ groups thus guarantees spin-degeneracy of the $\boldsymbol\Gamma$-point. On the other hand,  lifted spin-degeneracies in the rest of the Brillouin zone, including other time-reversal invariant momenta, are not generally excluded in altermagnets. We have  already derived above that the bands are symmetric with respect to the inversion of {\bf k}. Moreover, the collinearity, protected by the spin-only group symmetries, implies that spin is a good quantum number and spins have a common ${\bf k}$-independent quantization axis across the non-relativistic band structure. These basic spin-momentum locking characteristics of altermagnets are in striking contrast to the relativistic band structures of non-magnetic crystals. The latter systems have all time-reversal invariant momenta spin-degenerate, and the spin-momentum locking due to the broken real-space inversion symmetry is odd in ${\bf k}$, and has a form of a continuously varying spin-texture in the momentum space \cite{Winkler2003}. 

Other prominent spin-momentum locking features in altermagnets are protected by the specific $[{C}_2\parallel A] \, [E\parallel{\bf H}]$ symmetries present in the given ${\bf R}_s^{\rm III}$ group. For example, a symmetry $[{C}_2\parallel M_{c}]$, where $c$ is the axis perpendicular to the $a-b$ mirror plane, defines a spin-degenerate $k_a-k_b$ nodal plane at $k_c=0$, or other $k_c$ separated from $M_{c}k_c=-k_c$ by a reciprocal lattice vector.  This is because $[{C}_2\parallel M_{c}]$ transforms a wavevector from this plane on itself, or on an equivalent crystal momentum separated by the reciprocal lattice vector, while spin is reversed. Similarly, a $[{C}_2\parallel C_{n,c}]$ symmetry, where  $C_{n,c}$ is an $n$-fold rotation symmetry around the $c$-axis, imposes a spin-degenerate nodal-line parallel to the $k_c$-axis for wavevectors  with $k_a=k_b=0$, or other $k_{a(b)}$ separated from $C_{n,c}k_{a(b)}$ by a reciprocal lattice vector. We note that the high symmetry planes or lines are typically of main focus when assessing the electronic structures. This may explain why, apart from the omission by the conventional magnetic groups,  altermagnets remained unnoticed during the decades of  DFT and experimental studies of band structures.

Each of the 10 spin Laue groups of altermagnets can be assigned a characteristic spin winding number. In Tab.~1 we show the spin-momentum locking with the corresponding spin winding number  depicted on top of model ${\bf k}\cdot{\bf p}$ Hamiltonian  bands (see Supplementary Sec.~II). We obtain either planar (cf. relativistic Rashba or Dresselhaus spin-texture) or bulk (cf. relativistic Weyl spin-texture) non-relativistic spin-momentum locking, with the characteristic  spin winding number $W=2$, 4, or 6, when integrated over a surface near and enclosing the $\boldsymbol\Gamma$-point. (Further from the $\boldsymbol\Gamma$-point, additional band-crossings can result in higher even-integer winding numbers.) The planar altermagnetic spin-momentum locking is relevant for (quasi)two-dimensional and three-dimensional crystals, while the bulk spin-momentum locking only for three-dimensional crystals. We note that the earlier reported materials\cite{Lopez-Moreno2012,Noda2016,Smejkal2020,Ahn2019,Hayami2019,Naka2019,Yuan2020,Feng2020a,Hayami2020,Reichlova2020,Yuan2021a,Egorov2021,Mazin2021,Egorov2021a,Gonzalez-Hernandez2021,Naka2021,Bose2021,Bai2021,Smejkal2021,Shao2021} FeF$_2$, MnO$_2$, RuO$_2$,  $\kappa$-Cl, MnF$_2$, Mn$_5$Si$_3$, LaMnO$_3$,  FeSb$_2$, CaCrO$_3$, referred to as unconventional spin-split antiferromagnets in these studies,  all correspond   to the altermagnetic class with the characteristic planar spin winding number $W=2$. 

A complementary orbital-harmonic representation\cite{Hayami2020} allows us to identify a $d$-wave, $g$-wave, or $i$-wave form of the anisotropic spin-momentum locking near the $\boldsymbol\Gamma$-point.  The list of characteristic (not exclusive) anisotropic crystal harmonics, given in Supplementary Sec.~II, facilitated the derivation of the model ${\bf k}\cdot{\bf p}$ Hamiltonian bands of altermagnets shown in Tab.~1.

\subsection*{Spin-splitting by local electric crystal field}
We now demonstrate the spin-momentum locking protected by spin-group symmetries, and the microscopic spin-splitting mechanism due to the local electric crystal field, on an altermagnetic spin arrangement on the crystal of KRu$_4$O$_8$. The real-space crystal structure of KRu$_4$O$_8$,  as reported in earlier studies \cite{Kobayashi2009,Toriyama2011}, is schematically illustrated in Fig. 2a. The symmetry of the lattice is body-centred tetragonal (crystallographic space group $I4/m$). Purple and  cyan color in Fig. 2a represent the collinear antiparallel spin arrangement on the crystal.  In addition, the A and B symbols label the real-space sublattices corresponding to the opposite spins in the altermagnetic phase. The A and B real-space sublattices are strongly anisotropic, and related by a mutual planar rotation by  90$^\circ$ ($C_{4z}$). Correspondingly, the non-trivial spin Laue group describing the spin-momentum locking  in the altermagnetic phase is $2_{4/}1_m$. According to Eq.~(\ref{Rsstar}), it can be decomposed as,
\begin{equation}
2_{4/}1_m=[E\parallel 2/m] + [{C}_2\parallel C_{4z}] \, [E\parallel 2/m]. 
\label{group_KRuO}
\end{equation}

Fig.~2b shows the DFT calculation of the spin-momentum locking, protected by the spin-group symmetries, on top of two selected  KRu$_4$O$_8$ Fermi surface sheets.  (The band structure was obtained  using the DFT full-potential linearized augmented plane-wave code ELK within the local-spin-density approximation \cite{elk}.) In particular, the $[{C}_2\parallel C_{4z}]$ symmetry leads to three spin-degenerate nodal lines parallel to the $k_z$-axis, $[0,0,k_z]$, $[2,0,k_z]$, and  $[0,2,k_z]$, marked by grey points in Fig.~2b  (here the wavevectors are in units of $\pi$ divided by the lattice constant). The latter two correspond, for $k_z=0$, to  time-reversal invariant momenta ${\bf M}_1$ and ${\bf M}_2$. On the other hand, the spin-degeneracy is strongly lifted at time-reversal invariant momenta ${\bf X}$ and ${\bf Y}$, corresponding to the directions of the  real-space anisotropy axes of the two sublattices. Consistently, the little crystallographic Laue group at the ${\bf X}$ and ${\bf Y}$ wavevectors is $2/m$, which  coincides with the halving subgroup of same-spin-sublattice transformations. The resulting  spin winding number is $W=2$. The spin-momentum locking is planar, reflecting the real-space planar mutual rotations of the crystal-anisotropies of the opposite-spin sublattices. 

We now move on to the demonstration of the electric crystal-field mechanism of the spin-splitting, which we compare to the more conventional magnetic mechanism. The analysis is presented in Figs.~2c-g. Energy bands in the non-magnetic and altermagnetic phase are shown in Figs.~2c,d. The high-energy band around 0.9~eV in the depicted portion of the Brillouin zone is two-fold spin-degenerate  in the non-magnetic phase (upper part of Fig.~2c). The magnetic component of the internal electro-magnetic crystal potential in the altermagnetic phase generates an anisotropic {\bf k}-dependent spin splitting, as shown in the upper part  of Fig.~2d where the purple and cyan color correspond to opposite spin states. The sign of the spin splitting alternates, following the symmetries of the spin group. This type of the spin-splitting belongs to a family,   generally referred to as  internal magnetic-interaction mechanisms \cite{Yuan2020,Ahn2019,Hayami2020,Smejkal2021}. Note that another example, considered earlier in the literature, is a Pomeranchuk Fermi liquid instability in an anisotropic ($d$-wave) spin-triplet channel \cite{Ahn2019}. In contrast to this correlated physics interpretation, we identify the spin splitting  without including correlation effects beyond the local-spin-density approximation.

The other bands of KRu$_4$O$_8$, for energies near the Fermi level, also show a spin splitting within the DFT band-structure theory. However, here the microscopic origin is in the electric crystal field, i.e., it is fundamentally distinct from the internal magnetic-interaction mechanisms.  In the non-magnetic phase, we observe in the lower part of Fig.~2c a couple of two-fold spin-degenerate bands whose mutual splitting (highlighted by grey shading) by the electric crystal field is {\bf k}-dependent, merging at the  four-fold degenerate $\boldsymbol\Gamma$, ${\bf M}_1$ and ${\bf M}_2$ points. Remarkably, the {\bf k}-dependent spin-splitting in the altermagnetic phase in the lower part of Fig.~2d (highlighted again by grey shading) copies the {\bf k}-dependent band splitting by the electric crystal field in the non-magnetic phase.  Its microscopic explanation is provided in Figs.~2f,g. 

We start the discussion of Fig.~2f from the projections of the  bands on the sublattice A and B in the non-magnetic phase  (black lines). The projections are dominated by Ru $d_{xz}$ and $d_{yz}$ orbitals. This is in agreement with earlier report\cite{Toriyama2011}, which showed the presence of Ru $t_{2g}$ orbitals near the Fermi level. At the $\boldsymbol\Gamma$-point, the A and B  projected bands are degenerate, which is consistent with the octahedral environment with the tetragonal symmetry \cite{Khomskii2014}. Including spin, the $\boldsymbol\Gamma$-point is then four-fold degenerate in the non-magnetic phase. 

The band whose dominant weight is on sublattice A  is strongly anisotropic with respect to {\bf k} when moving towards the {\bf X} and {\bf Y} points (left panel of Fig.~2f). The same applies to the sublattice B   band, however, the sense of the anisotropy reverses (right panel of Fig.~2f). The band anisotropies  reflect the strong local crystalline anisotropy, conspiring with the favorable symmetry of the involved orbitals. The local crystalline anisotropy is described by the  elements of the halving subgroup $2/m$, which interchange atoms within the same  sublattice. On the other hand, the reversed sense of the anisotropies of the sublattice A and B  bands reflects the real-space rotation symmetry $C_{4z}$, which interchanges atoms between sublattice A and B. By adding up the A and B projections we obtain the bands shown in Fig.~2c. They progressively split by the electric crystal field  when the {\bf k}-vector moves  from the  $\boldsymbol\Gamma$-point towards, e.g., the {\bf X} point, with the lower band dominated by one sublattice and the upper band by the other sublattice. Along the $\boldsymbol\Gamma - {\bf Y}$ line, the sublattice indices of the lower and upper bands switch places. 

The bands in the altermagnetic phase, projected again on the sublattice A and B, are also plotted in Fig.~2f. As in Fig.~2d, the purple and cyan color correspond to opposite spins. We see that for bands with dominant weight on sublattice A, spin states  shown in purple move up in energy while the opposite spin states shown in cyan move down (left panel of Fig.~2f). The magnetic component of the internal crystal potential in the altermagnetic phase generates in this case a splitting (highlighted by light-blue shading) which is nearly  {\bf k}-independent. This scenario is fundamentally distinct from the strongly {\bf k}-dependent magnetic splitting of the high-energy band, shown in Fig.~2e. It is reminiscent of ferromagnets. However, unlike the common ferromagnetic case,  the nearly {\bf k}-independent magnetic splitting reverses sign in our altermagnet for the  sublattice B bands (right panel of Fig.~2f). This locality, in which band states near the Fermi level with one spin have a dominant weight on one sublattice, is again distinct from the delocalized nature of  spin states in the high-energy bands shown in Fig.~2e. It also starkly contrasts with the conventional mechanisms of the ferromagnetic splitting of band spin-states experiencing  the global magnetization, or the spin-orbit splitting due to the global electric inversion asymmetry. An additional illustration of the locality is shown in Fig.~2g where we plot  the real-space DFT spin-density  around the Ru atom in sublattice A and B. Consistent with the  spin Laue group symmetry and the dominant $d_{xz}$ and $d_{yz}$ orbitals near the Fermi level, the opposite-spin local densities in the two sublattices are highly anisotropic, with the mutually rotated real-space anisotropy axes. 

Adding up the A and B sublattice projections of Fig.~2f then explains the formation of two pairs of spin-split bands seen in Fig.~2d. The mutual magnetic splitting between the two pairs is nearly {\bf k}-independent, while the spin splitting within each pair is a {\bf k}-dependent copy of  the  band splitting by the local anisotropic electric crystal field in the non-magnetic phase (Fig.~2c). It also explains that the two pairs have opposite sign of the spin splitting and that, within each pair, the spin-splitting sign is opposite when moving from the $\boldsymbol\Gamma$-point point towards the {\bf X} or {\bf Y} points. We see from Figs.~2d and 2f, that even if the nearly {\bf k}-independent magnetic  splitting were small, the local  electric crystal field  would still determine the splitting between the two nearest bands with opposite spin in the altermagnetic phase at {\bf k}-vectors sufficiently close to the $\boldsymbol\Gamma$-point. This is a consequence of the nearly {\bf k}-independent magnetic band-splitting, and of the spin-degeneracy of the $\boldsymbol\Gamma$-point in altermagnets. 

In the studied KRu$_4$O$_8$ altermagnet, the spin-splitting originating from this extraordinary electric crystal-field mechanism reaches a 300~meV scale. In Supplementary Sec.~III and Fig.~S1 we show that in altermagnetic RuO$_2$, a spin splitting reaching a 1~eV scale\cite{Smejkal2020,Ahn2019,Smejkal2021} is also due to the electric crystal-field mechanism. These altermagnetic spin-splitting magnitudes are comparable to spin splittings in ferromagnets but, unlike ferromagnets, are accompanied by a zero net magnetization. They also illustrate that spin splittings in altermagnets can exceed by an order  of magnitude the record relativistic spin-orbit splittings in bulk crystals with heavy elements \cite{Ishizaka2011}. Moreover, unlike the spin-orbit split bands, altermagnetism preserves a common {\bf k}-independent spin quantization axis.

Finally, we emphasize that relativistic DFT calculations in RuO$_2$ and KRu$_4$O$_8$, presented in Supplementary Sec.~III,IV and Fig.~S1,S2, show the expected weak effect of the spin-orbit coupling on the bands.  This highlights that the apparent prominent features of the relativistic bands, including the spin-momentum locking characteristics and the electric crystal-field mechanism of the spin-splitting,   still reflect the non-relativistic spin-group symmetries. In contrast, these prominent symmetries are omitted by the relativistic magnetic groups of RuO$_2$ and KRu$_4$O$_8$. In general, as also illustrated in  Sec.~III,IV and Fig.~S1,S2, only the spin-group formalism facilitates the sublattice coset decomposition into transformations which interchange atoms between same-spin and opposite-spin sublattices, which plays the central role in the understanding of altermagnetism.

\subsection*{Candidate materials}

Tab.~1  lists selected candidate altermagnets.  In Fig.~3 we highlight CrSb, a metal with the N\'eel temperature of 705~K \cite{Park2020a}. As shown in Fig.~3a, it crystallizes in the hexagonal NiAs-type structure (crystal space group $P6_3/mmc$) \cite{Yuan2020a,Park2020a}. The collinear antiparallel spin arrangement corresponds to the non-trivial spin Laue group $2_{6/}2_m2_m1_m$ ($[E\parallel \bar{3}m] +[{C}_2\parallel C_{6z}] \, [E\parallel \bar{3}m]$). It contains the  $[{C}_2\parallel{M}_{z}]$ symmetry, which makes the spin-momentum locking bulk-like. Additional mirror planes orthogonal to the three hexagonal crystal axes, combined with the spin rotation, imply that the spin winding number is $W=4$. This is confirmed by the DFT calculations in Fig.~3b. 

CrSb has a more complex band structure than KRu$_4$O$_8$, as shown in Fig.~3c. Nevertheless, we can still trace a pair of  bands with opposite spin (highlighted by grey shading in Fig.~3c) which are degenerate at $\boldsymbol\Gamma$, ${\bf L}_1$ and ${\bf L}_2$ points, and split  when moving away from these high symmetry points. The spin splitting is as high as 1.2~eV. We also note that altermagnetic CrSb hosts an extraordinary spin-polarized quasiparticle which is four-fold degenerate at the $\boldsymbol\Gamma$-point and spin split away from the $\boldsymbol\Gamma$-point. In the Supplementary Sec.~V and Figs.~S3,S4  we discuss additional altermagnetic candidates  among insulators, semiconductors and metals, and give an example illustrating the inversion symmetry of the altermagnetic bands even when the crystal is inversion asymmetric. 

The variety of altermagnetic material types brings us back to  the representative altermagnetic structure shown in Fig.~1. It corresponds to the parent cuprate La$_2$CuO$_4$ of a high-temperature superconductor  \cite{Lane2018}. The band structure for the collinear antiparallel spin arrangement on this crystal falls into the altermagnetic non-trivial spin Laue group $2_m2_m1_m$ ($[E\parallel 2/m] + [{C}_2\parallel C_{2y}] \,[E\parallel 2/m]$).  The symmetry element $[{C}_2\parallel C_{2y}]$ generates a planar  spin-momentum locking with a characteristic spin winding number $W=2$.  Remarkably, according to our spin-group theory, the energy bands of La$_2$CuO$_4$  are altermagnetically spin split. This is confirmed by the DFT calculations in Supplementary Fig.~S5, and is in contrast with the conventional perception of spin-degenerate bands in La$_2$CuO$_4$  \cite{Lane2018}. The omission of the spin-splitting physics in earlier electronic-structure studies of cuprates could be explained by the focus on high-symmetry lines or planes, such as the $k_z=0$ plane\cite{Sobota2021}, where the states are spin-degenerate (see Supplementary Fig.~S5). Our work thus  brings a new element into the research of the coexistence and interplay of magnetic and superconducting quantum orders \cite{Si2016}.

\subsection*{Discussion}
Our extraordinary spin-splitting mechanism by the local electric crystal-field in the Ru-oxide altermagnets provides a unifying microscopic picture of recent intriguing theoretical and experimental observations of spin-transport anomalies in the magnetically compensated RuO$_2$, namely of the large spontaneous Hall, charge-spin conversion and spin-torque phenomena \cite{Smejkal2021b,Smejkal2020,Feng2020a,Gonzalez-Hernandez2021,Bose2021,Bai2021,Smejkal2021,Shao2021}. We also note that our identification of altermagnetism in chalcogenide CoNb$_3$S$_6$, perovskite CaMnO$_3$, or the cuprate La$_2$CuO$_4$ crystals sheds new light on puzzling magneto-transport anomalies found in literature on these materials \cite{Tenasini2020,Vistoli2019,Wu2017c}. 

The diversity of the altermagnetic material types illustrates the relevance of this magnetic phase for a range of condensed-matter physics fields. Apart from superconductors, unexplored connections might exist between altermagnetism and topological insulators and semimetals. In this context we note that, on one hand, symmetry prohibits a realization of altermagnetism in one-dimensional chains;  collinear antiferromagnetic spin arrangements on one-dimensional chains have the $[{C}_2\parallel \bar{E}]$ (and possibly also $[{C}_2\parallel {\bf t}]$) symmetry and, therefore, have Kramers spin-degenerate bands. On the other hand, we have identified altermagnetic candidates among quasi-one-dimensional, quasi-two-dimensional, and three-dimensional insulators and metals. This opens the possibility of searching for unconventional spin-polarized fermion quasiparticles (cf. CrSb), topological-insulators and topological-semimetals, including Chern insulators with the quantized Hall effect in high-temperature systems with vanishing internal or external magnetic dipole.

We expect that the altermagnetic spin-momentum locking, with the extraordinary spin-splitting mechanism by the local electric crystal field, will play a special role in the search for new quantum phases, complementing the widely explored relativistic, magnetic, and many-body correlation physics. This applies to both basic science and technology contexts. Regarding the latter,
our spin-group classification of spin-momentum locking for collinear spin arrangements, together with DFT calculations in representative crystals, theoretically underpins altermagnetism as a favourable magnetic phase for research and applications of charge-spin conversion devices, including magnetic memories.  Among the candidate altermagnetic materials, we find high transition-temperature altermagnets composed of abundant light elements, which have a large energy separation of opposite spin states. 
Moreover, unlike ferromagnets, they eliminate stray fields and add insensitivity to external magnetic field perturbations. 
When comparing to the 
relativistic non-magnetic spin-texture phases, these  share with altermagnets  the zero net dipole-moment. However, large relativistic spin splittings require rare heavy elements. In addition, 
the relativistic phases suffer from spin decoherence even for small-angle elastic scattering off common isotropic impurities. We have illustrated that this obstacle is diminished in altermagnets by the collinearity of spins, and by the possibility of a large {\bf k}-vector separation in the Brillouin zone of the equal-energy eigenstates with the opposite spin.

\newpage

\begin{figure}[h!]
\begin{center}
\hspace*{-0cm}\epsfig{width=1\columnwidth,angle=0,file=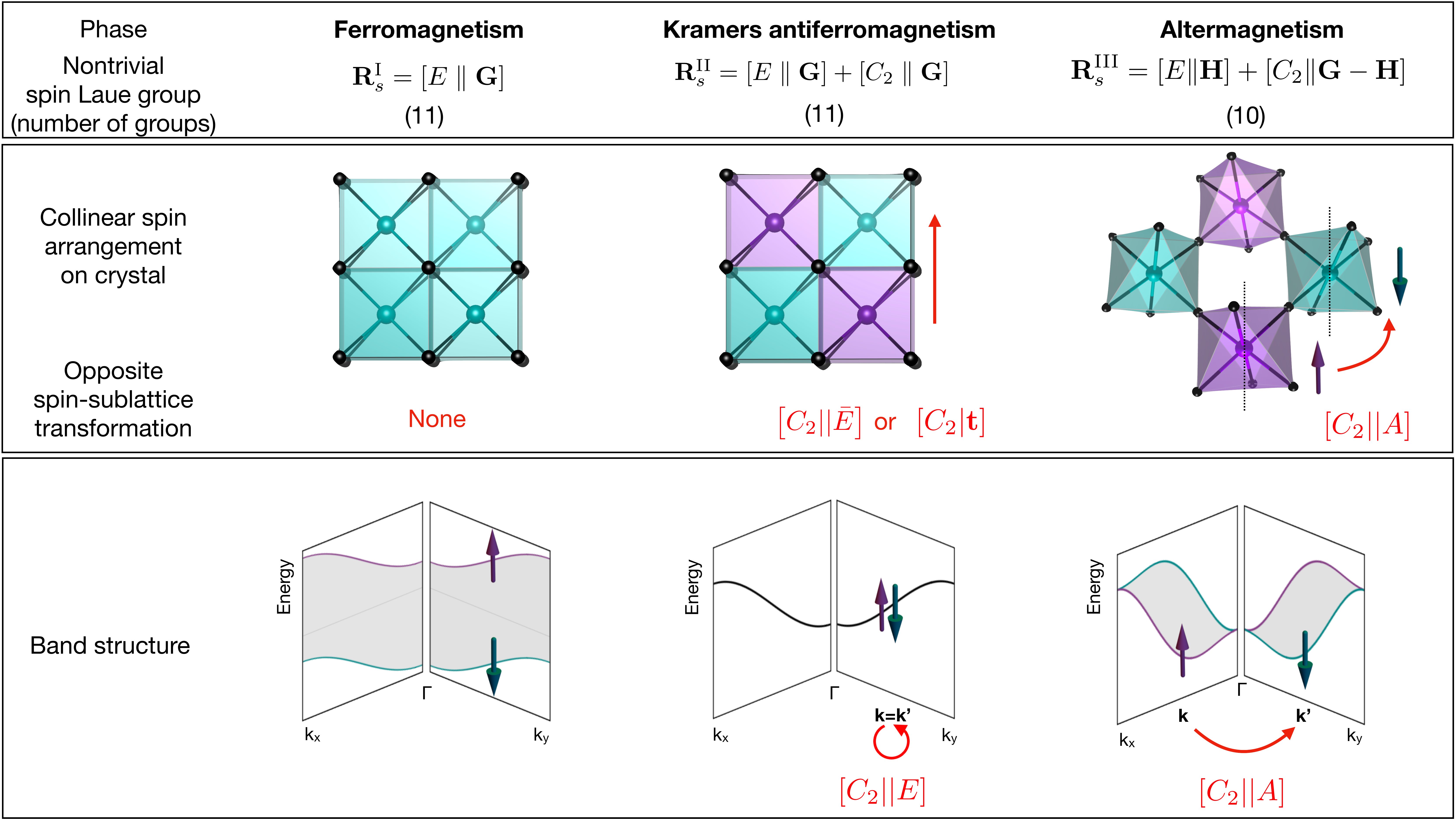}
\end{center}
\vspace{0cm}
\caption{\footnotesize
{\bf{Illustration (in columns) of ferromagnetic, Kramers spin-degenerate antiferromagnetic, and altermagnetic phases. }} 
{\bf First row}: Non-trivial spin Laue group structure for the given magnetic phase with the number of different groups in brackets. {\bf Second row}: Illustrative collinear spin arrangements on crystals. Opposite spin directions are depicted by purple and cyan color.  Instead of marking the spin arrangement by arrows, we use a two-color representation to highlight that the overall spin axis orientation is not related to the real space coordinates for the non-relativistic spin-group symmetries. Ferromagnetic and Kramers antiferromagnetic crystals correspond to FeRh, the altermagnetic crystal to La$_2$CuO$_4$. Red arrow and its label highlights opposite-spin-sublattice transformation symmetries  which can generate Kramers spin-degenerate antiferromagnetism (real-space translation or inversion) and altermagnetism (real-space rotation). The collinear antiparallel spin arrangement in the altermagnet is further highlighted by the  purple and cyan arrows next to the opposite spin sublattices. {\bf Third row}: Band-structure cartoons 
 show ferromagnetically spin-split  bands (opposite spin states depicted by purple and cyan color), a  Kramers spin-degenerate antiferromagnetic band, and altermagnetically spin-split bands. The opposite-spin-sublattice transformation of the spin Laue group which maps the same-energy eigenstates with opposite spins on the same {\bf k}-vector in Kramers antiferromagnets and on different  {\bf k}-vectors in altermagnets is again highlighted. 
} 
\label{fig1}
\end{figure}

\pagebreak

\begin{table}[h!]
\begin{center}
\hspace*{-0cm}\epsfig{width=1\columnwidth,angle=0,file=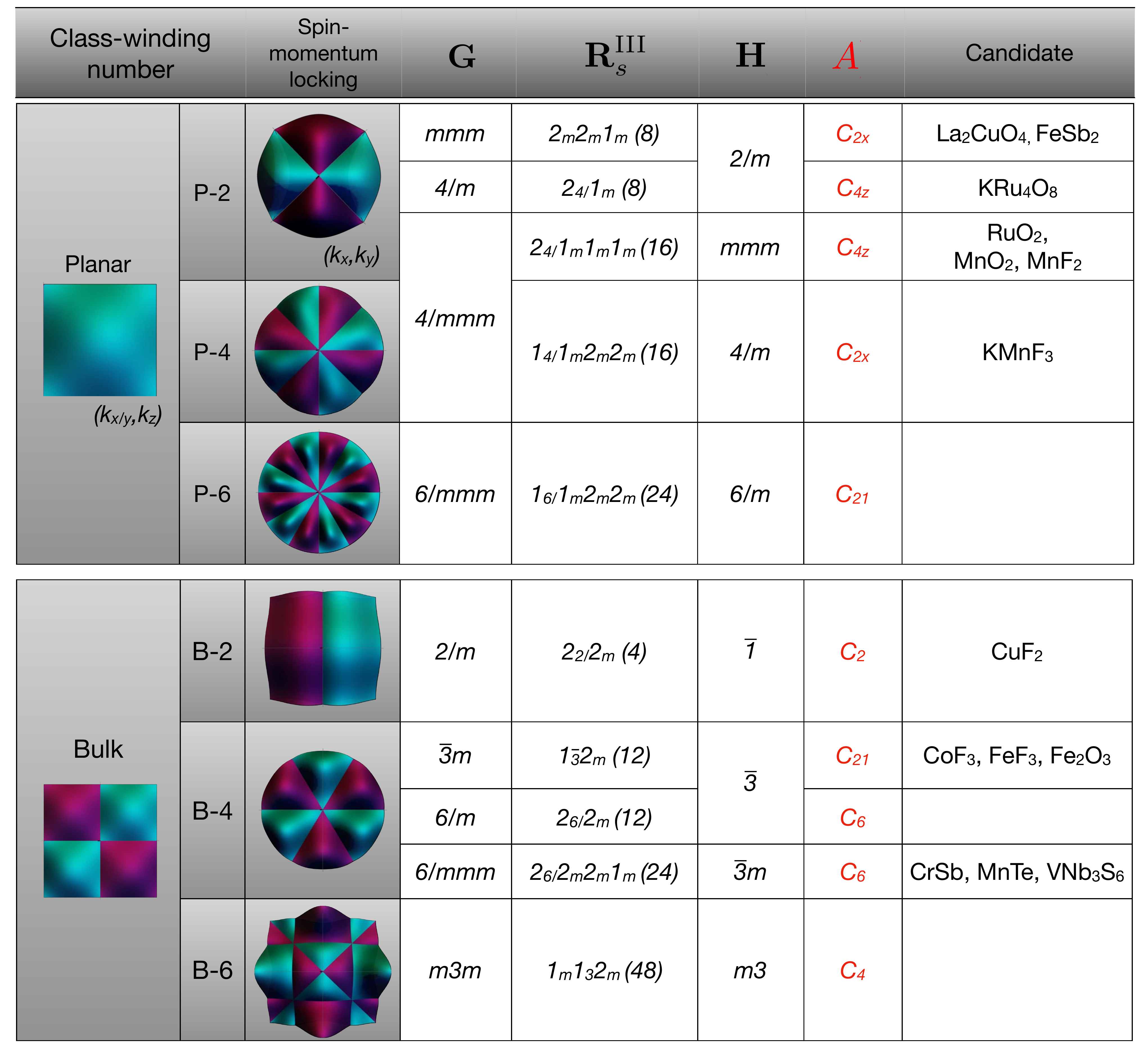}
\end{center}
\vspace{-1cm}
\caption{\footnotesize {\bf Classification of altermagnetic spin-momentum locking protected by spin-group symmetries, and material candidates.} The columns describe, respectively, planar (P) or bulk (B)  spin-momentum locking 
with the corresponding spin winding number,
 the crystallographic Laue group {\bf G}, the halving subgroup {\bf H} of symmetry elements which interchange atoms between same-spin sublattices,   a generator $A$ of symmetry elements which interchange atoms between opposite-spin sublattices, the non-trivial spin Laue group $R_s^{\rm III}$ (in the bracket we list the number of symmetry elements),  and altermagnetic material candidates. The model ${\bf k}\cdot{\bf p}$ Hamiltonian bands on which we illustrate the spin-momentum locking character and spin winding number are described in Supplementary Sec.~II. References describing the materials are in the main text and Supplementary Sec.~III-V.} 
\end{table}

\pagebreak

\begin{figure}[h!]
\begin{center}
\hspace*{-0cm}\epsfig{width=1\columnwidth,angle=0,file=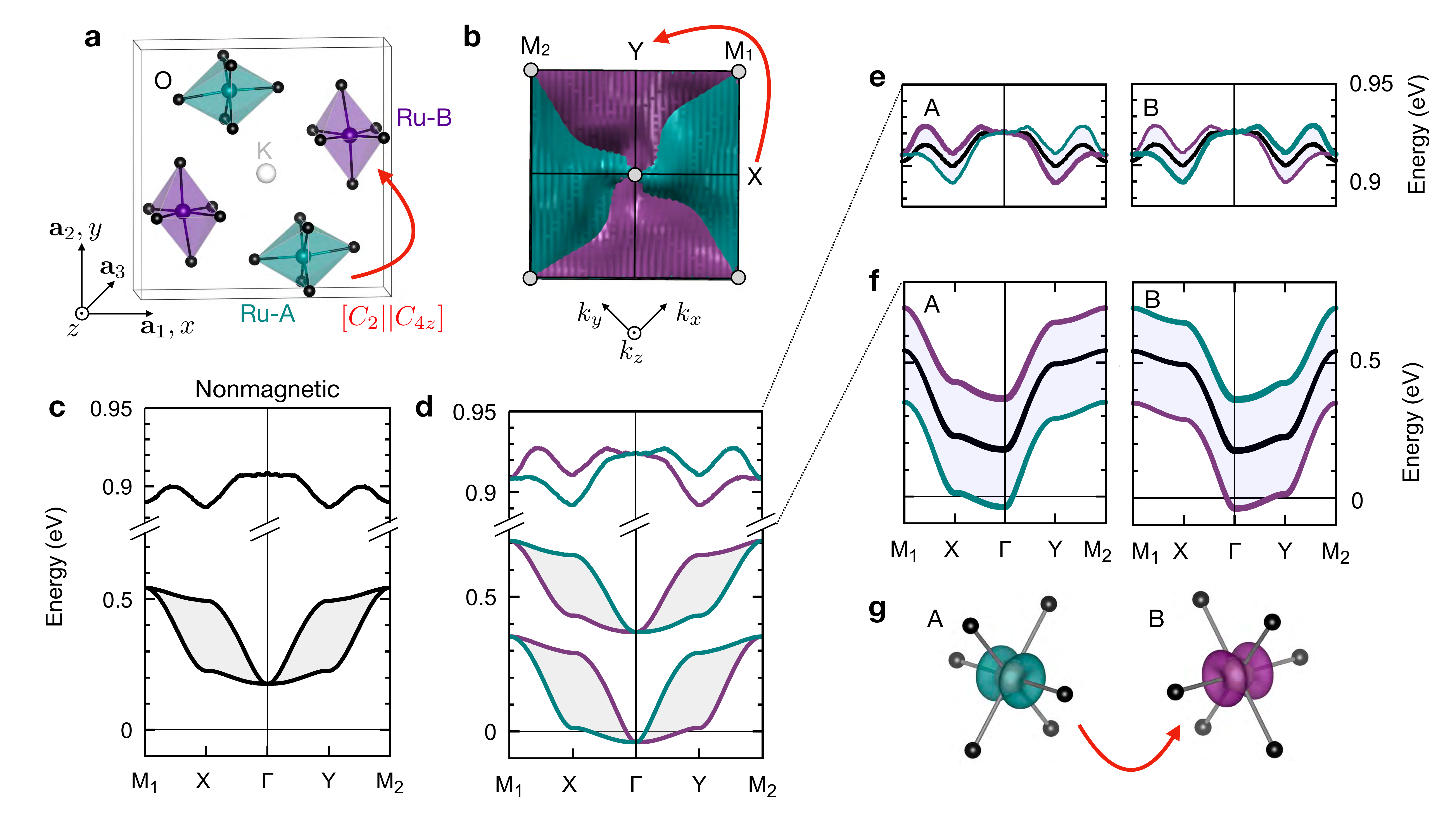}
\end{center}
\caption{{\bf{Spin splitting by local electric crystal field in altermagnetic KRu$_4$O$_8$.}} {\bf a}, Schematic spin arrangement on the KRu$_4$O$_8$ crystal with opposite spin directions depicted by purple and cyan color. Red arrow and its label highlights the opposite-spin-sublattice transformation, containing a real-space four-fold  rotation. {\bf b}, Calculated spin-momentum locking with the spin winding number $W=2$ on top of two  DFT Fermi surface sheets. {\bf c,d}, DFT band structure of the non-magnetic and altermagnetic phase, respectively. Grey shading highlights the  {\bf k}-dependent splitting  by the local electric crystal field. {\bf e,f}, Projection of bands on the sublattice A and B  in the non-magnetic (black) and altermagnetic (purple and cyan) phase for the upper bands and lower bands, respectively. Color shading in panel f highlights the nearly  {\bf k}-independent magnetic splitting of the lower bands, and its opposite sign for the sublattice A and B bands. {\bf g},  Real-space DFT spin density  around the Ru atom in sublattice A and B.
}
\label{fig3}
\end{figure}

\pagebreak

\begin{figure}[h!]
\begin{center}
\hspace*{-0cm}\epsfig{width=1\columnwidth,angle=0,file=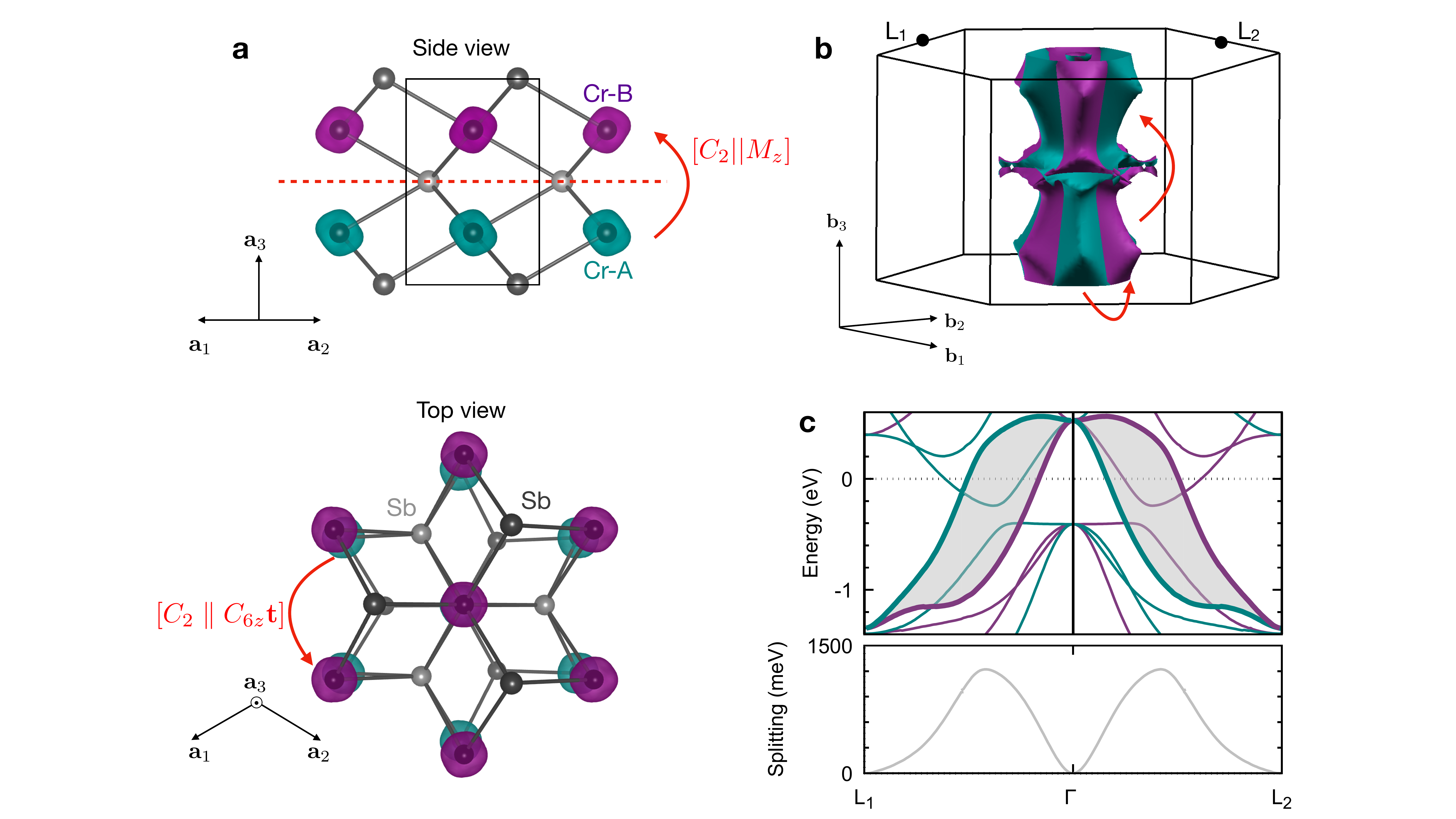}
\end{center}
\caption{{\bf{Metallic high N\'eel temperature altermagnet CrSb.}} {\bf a}, Schematic crystal structure with  DFT spin densities. Cr sublattices and the respective magnetization densities with opposite orientation of the magnetic moment are depicted by purple and cyan color. Red arrow and its label highlights the opposite-spin-sublattice transformation, containing a  real-space mirror or six-fold rotation. {\bf b}, Calculated spin-momentum locking with the bulk spin winding number $W=4$ on top of two selected DFT  Fermi surface sheets.  {\bf c},  DFT band structure in the altermagnetic phase.  Wavevector dependence of the spin splitting between the bands highlighted  by the grey shading is plotted in the lower panel.} 
\label{figCrSb}
\end{figure}

\newpage

\begin{center}
{\bf SUPPLEMENTARY INFORMATION}
\end{center}

\section{Details of the derivation of spin group categorization of non-relativistic collinear magnetism}  
The first type of the non-trivial spin Laue groups is obtained from the  group of spin-space transformations ${\bf S}_1=\{E\}$ whose only possible coset decomposition contains one coset, which is ${\bf S}_1$ itself. The corresponding isomorphic one-coset decomposition of a crystallographic Laue group ${\bf G}$ is, therefore, again ${\bf G}$ itself. As a result, the first type of non-trivial spin Laue groups is given by ${\bf R}_s^{\rm I}=[E\parallel{\bf G}]$. 

The  second type of the non-trivial spin Laue groups is obtained from the one-coset decomposition of ${\bf S}_2=\{E,C_2\}$, i.e.,   ${\bf S}_2$ itself. The corresponding non-trivial spin Laue groups are then given by ${\bf R}_s^{\rm II}=[E\parallel{\bf G}]+[{C}_2\parallel{\bf G}]$.   Below we will show that the ${\bf R}_s^{\rm II}$ groups also describe Kramers spin-degenerate collinear antiferromagnetism in crystals with the opposite-spin-sublattice transformation symmetry  $[{C}_2\parallel \bar{E}]$, where $\bar{E}$ on the right side of the double vertical bar is the real-space inversion. 

Since ${\bf S}_2$ has  two elements, its only additional coset decomposition contains two cosets, ${\bf S}_2=\{E\} + C_2\{E\}$. The corresponding isomorphic two-coset decomposition of the crystallographic Laue group is given by, ${\bf G}={\bf H}+ A{\bf H}={\bf H}+({\bf G}-{\bf H})$, where  {\bf H}  is a halving subgroup of {\bf G} and $A$ is any chosen element of ${\bf G-H}$. The non-trivial spin Laue groups obtained from this coset decomposition take a form,
\begin{equation}
{\bf R}_s^{\ast}=[E\parallel{\bf H}]+[{C}_2\parallel A] \, [E\parallel{\bf H}]=[E\parallel{\bf H}]+[{C}_2\parallel{\bf G-H}].
\label{Rsstar}
\end{equation}
We see from Eq.~(\ref{Rsstar}) that for ${\bf R}_s^{\ast}$, {\bf G} is expressed as a sublattice coset decomposition, where  {\bf H} contains only the real-space transformations which interchange atoms between same-spin sublattices, and ${\bf G}-{\bf H}$ contains  only the real-space transformations which interchange atoms between opposite-spin sublattices.

Because of the symmetry $[\bar{C}_2\parallel {\cal T}]$  of the collinear spin arrangements, following from their spin-only group, and because $[\bar{C}_2\parallel {\cal T}]$ acts the same on $\epsilon(s,{\bf k})$ as $[E\parallel \bar{E}]$, the latter is also a symmetry of all non-relativistic collinear magnets. This turns Eq.~(\ref{Rsstar}) in two distinct types, depending on whether the real-space inversion is an element of ${\bf G-H}$, i.e. belongs to the opposite-spin-sublattice transformations, or is an element of ${\bf H}$, i.e. belongs to the same-spin-sublattice transformations. In the former case, $[E\parallel \bar{E}] [E\parallel{\bf H}]=[E\parallel{\bf G}]$, and ${\bf R}_s^{\ast}$ becomes ${\bf R}_s^{\rm II}$. This explains the above statement that the ${\bf R}_s^{\rm II}$ Kramers spin-degenerate antiferromagnetic phase is also obtained when the real-space inversion interchanges atoms between opposite-spin sublattices.

In the other case when the real-space inversion is an element of ${\bf H}$, $[E\parallel \bar{E}] [E\parallel{\bf H}]=[E\parallel{\bf H}]$, and we arrive at the remaining third distinct type of the non-trivial spin Laue groups, ${\bf R}_s^{\rm III}$. It is given by Eq.~(\ref{Rsstar}) with $A$ representing a real-space proper or improper rotation  which interchanges atoms between opposite-spin sublattices.

\section{Model ${\bf k}\cdot{\bf p}$ Hamiltonians of altermagnets}
The six model ${\bf k}\cdot{\bf p}$ Hamiltonian bands in Tab.~1 in the main text, representing planar ($P$) and bulk ($B$) altermagnetic spin-momentum locking with the spin winding number $W=2$, 4, and 6 around the  $\boldsymbol\Gamma$-point, are given by

\begin{eqnarray}
H_{P-2,d}&=&J k_{x}k_{y}, \\
H_{P-4,g}&=&J k_{x}k_{y}(k_{x}^{2}-k_{y}^{2}), \\
H_{P-6,i}&=&J  k_{x}  k_{y} \left( (\sqrt{3}k_{x})^{2} - k_{y}^{2}\right)\left( (\sqrt{3}k_{y})^{2} -  k_{x}^{2}\right),
\end{eqnarray}
which are independent of $k_{z}$, and by 
\begin{eqnarray}
H_{B-2,d}&=&J k_{z}k_{x}, \\
H_{B-4,g}&=&Jk_{z} k_{x}\left(k_{x}^{2}-\left(\sqrt{3} k_{y}\right)^{2}\right), \\
H_{B-6,i}&=&J\left(k_{x}^{2}-k_{y}^{2}\right)\left(k_{y}^{2}-k_{z}^{2}\right)\left(k_{z}^{2}-k_{x}^{2}\right).
\end{eqnarray}
which are $k_{z}$-dependent.
Here $J$ is the strength of the spin splitting and $d$, $g$, and $i$ refer to the $d$-wave, $g$-wave, and $i$-wave anisotropic crystal harmonic symmetry. The previously reported  materials and models\cite{Lopez-Moreno2012,Noda2016,Smejkal2020,Ahn2019,Hayami2019,Naka2019,Yuan2020,Feng2020a,Hayami2020,Reichlova2020,Yuan2021a,Egorov2021,Mazin2021,Egorov2021a,Gonzalez-Hernandez2021,Naka2021,Bose2021,Smejkal2021,Shao2021}
belong the the $H_{P-2,d}$ class.   

{\section{Spin splitting by electric crystal field and  comparison to relativistic magnetic groups and DFT calculations for ${\rm\bf RuO}_2$}

We start by illustrating the difference between the spin group and the magnetic group of RuO$_2$. In Supplementary Fig.~S\ref{RuO2}a we give the spin group of altermagnetic RuO$_2$,  $2_{4}/1_{m}1_{m}1_{m}$. We also explicitly indicate the generators $[E\parallel M_i]$, where $M_i$ are the three orthogonal mirror planes, of the same-spin-sublattice transformations (black), and the generator $[{C}_2\parallel C_{4z}]$ of the opposite-spin-sublattice transformations (red). 

In Supplementary Fig.~S\ref{RuO2}c we give the magnetic group  $m'm'm$ for the magnetic moment vectors along the [110]-axis shown in the panel by purple and cyan arrows. We include also its generators which are one mirror plane orthogonal to the magnetic moments (black) and two mirror planes parallel to the moments combined with time-inversion (blue). We highlight that all these symmetry elements are same-spin-sublattice transformations and that the magnetic group contains no opposite-spin-sublattice transformation elements. We also point out that two of the same-spin-sublattice mirror transformations are combined with the time-inversion, while the third one is not combined with the time-inversion. In general, magnetic group symmetries can have same-spin-sublattice transformations with or without time-inversion, and opposite-spin-sublattice transformations with or without time-inversion. This contrasts with the sublattice coset decomposition form of all altermagnetic ${\bf R}_s^{\rm III}$ spin groups, whose elements are divided into the same-spin-sublattice transformations $[E\parallel{\bf H}]$ and the opposite-spin-sublattice transformations $[{C}_2\parallel{\bf G-H}]$.

Supplementary Fig.~S\ref{RuO2}a also illustrates that, in general, the altermagnetic and ferromagnetic phases are always described by distinct non-relativistic spin groups. In contrast, Supplementary Fig.~S\ref{RuO2}c demonstrates that the same relativistic magnetic group can, in some cases, describe both the antiparallel and parallel order of the magnetic moments. This highlights that the magnetic groups cannot, in general,  discriminate between the two phases.

In Supplementary Figs.~S\ref{RuO2}b,d we compare DFT calculations of Fermi surface cuts at wavevector $k_z = 0$ calculated without and with relativistic spin-orbit coupling, respectively. Even for the heavy atoms  such as Ru, the spin-orbit coupling represents  only a small perturbative contribution. The apparent prominent features of the relativistic Fermi surface, including the four-fold spin-momentum locking characteristic,   still reflect the non-relativistic spin-group symmetries. In contrast, these prominent symmetries are omitted by the relativistic magnetic group.

In the main text, we demonstrate the electric crystal-field mechanism of the spin splitting in KRu$_4$O$_8$. Here we illustrate that this extraordinary mechanism is abundant in altermagnets with orbital degeneracies in the non-magnetic phase. In Supplementary Fig.~S\ref{RuO2}e we identify the mechanism  in altermagnetic RuO$_{\text{2}}$, in which earlier works  reported a large spin splitting on the eV scale\cite{Smejkal2020,Ahn2019}.   


\begin{figure}[h!]
\begin{center}
\hspace*{-0cm}\epsfig{width=1\columnwidth,angle=0,file=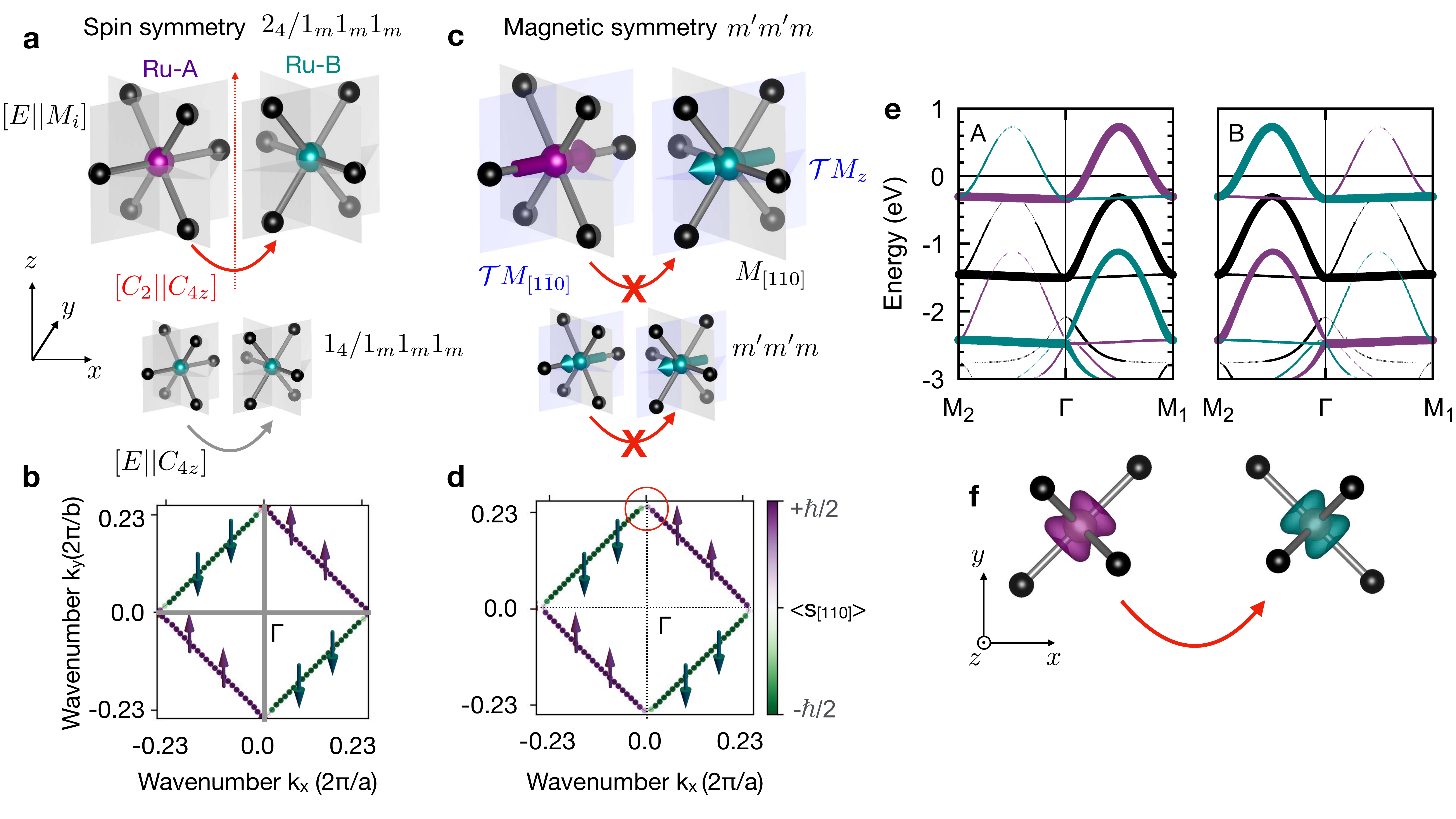}
\end{center}
\caption{{\bf Comparison of spin vs.  magnetic group and non-relativistic vs. relativistic DFT calculation, and identification of the spin splitting by local electric crystal field in  RuO$_{\text{2}}$.} 
{\bf a}, Schematic spin arrangement on the RuO$_{\text{2}}$ crystal with opposite spin directions depicted by purple and cyan color, and with the depicted non-relativistic spin group. Red arrow and its label highlights the generator  of opposite-spin-sublattice transformations, and the generators of the same-spin-sublattice transformations are also highlighted (in black). Bottom smaller image corresponds to the parallel spin-arrangement. {\bf c}, Schematic spin arrangement on the RuO$_{\text{2}}$ crystal with opposite spin directions and the crystallographic spin-axis orientation depicted by purple and cyan arrows, and with the depicted relativistic magnetic group and its generators. Bottom smaller image corresponds to the parallel spin-arrangement. The crossed red arrow highlights that the magnetic group contains no opposite-spin-sublattice transformation elements. {\bf b,d} Fermi surface cuts at wavevector $k_z = 0$ without and with relativistic spin-orbit coupling included in the DFT calculation, respectively. {\bf e}, Projection of the non-relativistic DFT bands on the sublattice A and B $m_l=+1$ orbitals   in the non-magnetic (black) and altermagnetic (purple and cyan) phase. {\bf f}, Real-space DFT spin density  around the Ru atom in sublattice A and B.
}
\label{RuO2}
\end{figure}

\newpage

\section{Comparison to relativistic magnetic groups and DFT calculations for ${\rm\bf KRu}_4{\rm\bf O}_8$}

Here we illustrate the difference between spin groups and magnetic groups on KRu$_4$O$_8$. In Supplementary Fig.~S\ref{KRu4O8}a we recall the spin group of altermagnetic KRu$_4$O$_8$,  $2_{4}/1_{m}$, and explicitly write the generators $[E\parallel \bar{E}]$ and $[E\parallel M_z]$ of the same-spin-sublattice transformations, and the generator $[{C}_2\parallel C_{4z}]$ of the  opposite-spin-sublattice transformations. In Supplementary Fig.~S\ref{KRu4O8}c we give the magnetic group  $2'/m'$ for the magnetic moments along the [100]-axis shown in the panel by purple and cyan arrows, and include also its generators which are the crystallographic space-inversion  $\bar{E}$ and the mirror symmetry $M_z$, combined with time-inversion. We highlight that both these symmetry elements are same-spin-sublattice transformations and that the magnetic group contains no opposite-spin-sublattice transformation elements. 

In Supplementary Figs.~S\ref{KRu4O8}b,d we compare  projections of bands on the sublattice A and B   without and with spin-orbit coupling included in the DFT calculation, respectively. We see that in both cases, band states with one dominant spin projection have a dominant weight on one sublattice. (Note that the small weight on the other sublattice seen in Supplementary Fig.~S\ref{KRu4O8}c was neglected when plotting Fig.~2f in the main text.) By comparing Supplementary Figs.~S\ref{KRu4O8}b,d we see that the main features of the energy bands, i.e., the local electric crystal-field mechanism of the spin splitting and the spin-momentum locking are captured by the spin-group symmetries.  Even for the heavy atoms  such as Ru, the spin-orbit coupling represents  only a small perturbative contribution. We observe the largest relativistic contributions around the $\boldsymbol\Gamma$-point, marked by the red circles. In this region the spin texture is influenced by the spin-orbit coupling  which leads to a small reduction  of the spin-projection along the N\'eel vector and a small relativistic spin splitting. The comparison of Supplementary Figs.~S\ref{KRu4O8}b,d  also demonstrates the dominant non-relativistic origin of the large spin splitting at the {\bf X} and {\bf Y} time-reversal invariant momenta.

\begin{figure}[h!]
\begin{center}
\hspace*{-0cm}\epsfig{width=1\columnwidth,angle=0,file=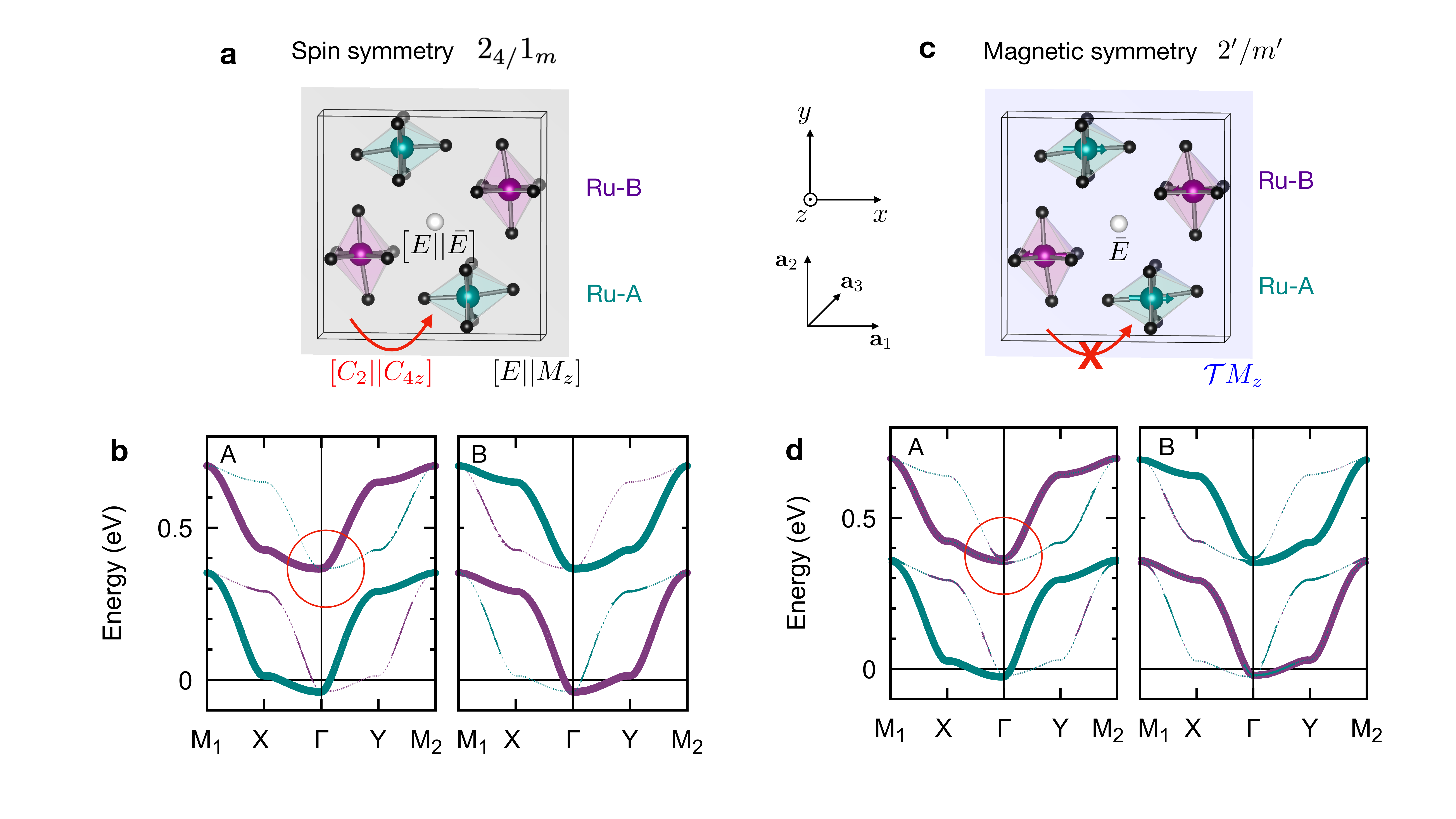}
\end{center}
\caption{{\bf Comparison of spin vs.  magnetic group and non-relativistic vs. relativistic DFT calculation for ${\rm\bf KRu}_4{\rm\bf O}_8$.} {\bf a}, Schematic spin arrangement on the KRu$_4$O$_8$ crystal with opposite spin directions depicted by purple and cyan color, and with the depicted non-relativistic spin group. Red arrow and its label highlights the generator of opposite-spin-sublattice transformations, and the generators of same-spin-sublattice transformations are also highlighted (in black). {\bf c}, Schematic spin arrangement on the KRu$_4$O$_8$ crystal with opposite spin directions and the crystallographic spin-axis orientation depicted by purple and cyan arrows, and with the depicted relativistic magnetic group and its generators. The crossed rad arrow highlights that the magnetic group contains no opposite-spin-sublattice transformation elements. {\bf b,d}, Projection of bands on the sublattice A and B without and with relativistic spin-orbit coupling included in the DFT calculation, respectively.} 
\label{KRu4O8}
\end{figure}

\newpage

\section{Other selected altermagnetic candidates}
In this section we show DFT calculations in the altermagnetic insulator CuF$_2$,  non-centrosymmetric altermagnet VNb$_3$S$_6$, and a parent cuprate altermagnet La$_2$CuO$_4$  of a high-temperature superconductor.  Beside these materials, Tab.~1 in the main text lists also  other insulating altermagnetic candidates, including high N\'eel temperature CoF$_3$ and FeF$_3$ \cite{Lee2018a},  and semiconducting MnTe \cite{Gonzalez-Hernandez2021}.  We also found altermagnetism in chalcogenite CoNb$_3$S$_6$ \cite{Tenasini2020} with bulk spin-winding number $W=4$, or in perovskite CaMnO$_3$ \cite{Vistoli2019} with planar spin-winding number $W=2$.

\subsection*{Altermagnetic insulator CuF$_2$}

In Supplementary Fig.~S\ref{CuF2} we present the symmetry analysis and DFT band-structure calculations of insulating CuF$_2$ with the N\'eel temperature 69~K \cite{Joenk1965} and with relatively light elements. Its crystal, shown in Fig.~S\ref{CuF2}a,  is a distorted rutile with a monoclinic structure (crystal space group $P2_1/$c) \cite{Zheng2012}. The altermagnetic non-trivial spin Laue group $2_{2/}2_m$ ($[E\parallel \bar{1}] +[C_2\parallel C_{2z}] \, [E\parallel \bar{1}]$) implies a spin winding number $W=2$. Unlike  KRu$_4$O$_8$, however, the  symmetry $[C_2\parallel M_z]$, which  complements the symmetry $[C_2\parallel C_{2z}]$, generates a bulk-like spin-momentum locking. This is confirmed by the DFT calculation presented in Fig.~S\ref{CuF2}b. We used the DFT+U method with $U=5$~eV and $J=1$~eV  to describe this Mott insulator\cite{Zheng2012}. The spin-resolved band structure is plotted in Fig.~S\ref{CuF2}c.  Similar to KRu$_4$O$_8$, the spin splitting is on the $\sim$100~meV scale in CuF$_2$. We note that the gap between the valence and conduction bands disappears for DFT with $U=J=0$. However, within a 15\% scatter, the altermagnetic spin splitting in the DFT band structure remains  the same as in the Mott insulating state obtained using DFT+U. This further underlines the robustness of altermagnetism, and of the spin splitting which originates from  the electro-magnetic crystal potential.

\newpage

\begin{figure}[h!]
\begin{center}
\hspace*{-0cm}\epsfig{width=1\columnwidth,angle=0,file=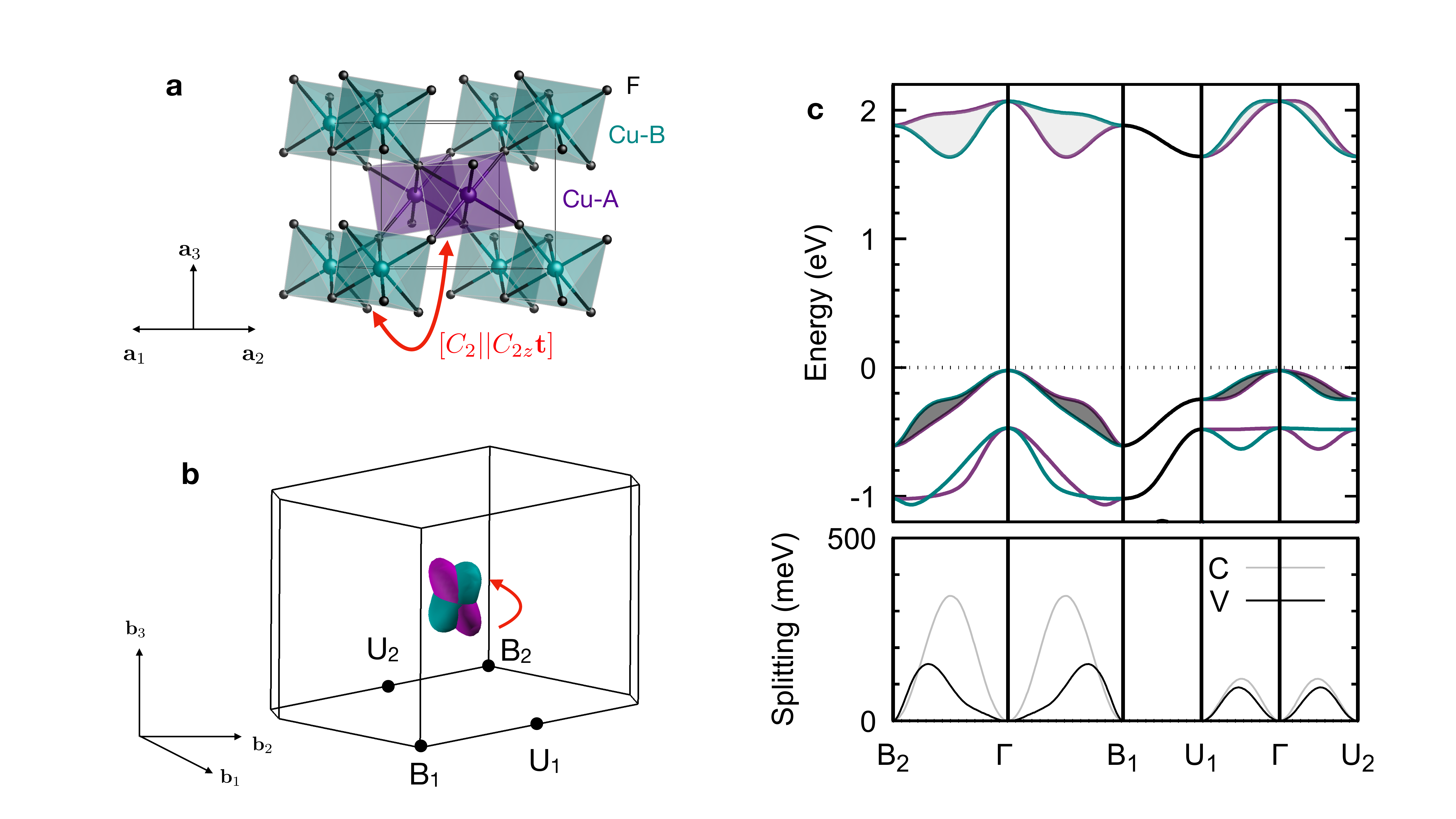}
\end{center}
\caption{{\bf{Altermagnetic insulator CuF$_2$.}} {\bf a}, Schematic spin arrangement on the CuF$_2$ crystal with opposite spin directions depicted by
purple and cyan color. Red arrow and its label highlights the opposite-spin-sublattice transformation, containing a real-space mirror or two-fold
rotation (combined with real-space translation). {\bf b}, Calculated spin-momentum locking with the bulk spin winding number $W=2$ on top of the DFT  Fermi surface for the Fermi level near the top of the valence band.  {\bf c},  DFT band structure in the altermagnetic phase. Wavevector dependence of the spin splitting between the
conduction bands highlighted by the light-grey shading and valence bands highlighted by the dark-grey shading is plotted in the lower panel by light and dark lines, respectively.} 
\label{CuF2}
\end{figure}

\newpage

\subsection*{Altermagnetic non-centrosymmetric VNb$_3$S$_6$}
In Supplementary Fig.~S\ref{VNb3S6} we present the symmetry analysis and DFT band-structure calculations of non-centrosymmetric  VNb$_3$S$_6$ with the N\'eel temperature 50~K \cite{Lu2020}. Its non-centrosymmetric crystal\cite{Lu2020}, shown in Fig.~S\ref{VNb3S6}a,  has the crystal space group $P6_322$. The altermagnetic non-trivial spin Laue group $2_{6/}2_m2_m1_m$ implies bulk spin winding number $W=4$.
\begin{figure}[h!]
\begin{center}
\hspace*{-0cm}\epsfig{width=1\columnwidth,angle=0,file=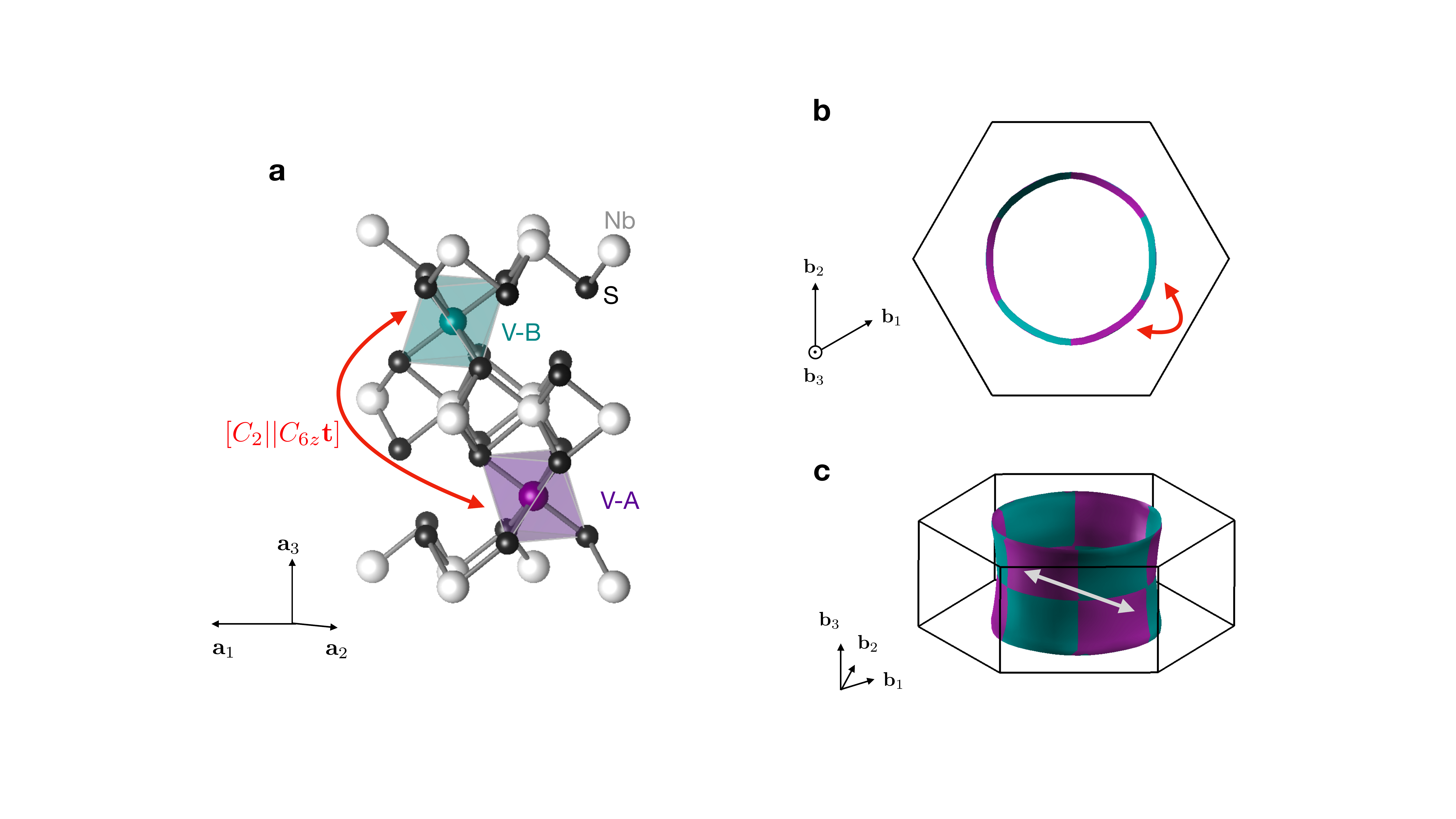}
\end{center}
\caption{{\bf{Altermagnetic non-centrosymmetric VNb$_3$S$_6$.}} {\bf a}, Schematic spin arrangement on the VNb$_3$S$_6$ crystal with opposite spin directions depicted by
purple and cyan color.  Red arrow and its label highlights the opposite-spin-sublattice transformation, containing a real-space six-fold
rotation. {\bf b,c}, Calculated  spin-momentum locking with the bulk spin winding number  $W=4$  on top of the DFT  Fermi surface. The  altermagnetic spin-momentum locking is symmetric with respect to the inversion of {\bf k}, despite the non-centrosymmetric crystal structure, as highlighted by the white double-arrow in panel c.  
}
\label{VNb3S6}
\end{figure}

\newpage

\subsection*{Altermagnetic parent cuprate  La$_2$CuO$_4$  of a high-temperature superconductor}

In Supplementary Fig.~S\ref{La2CuO4}a we show the crystal structure of La$_2$CuO$_4$ in the orthorhombic phase with tilted oxygen  octahedra. The spin-momentum locking in the altermagnetic phase is planar with the spin winding number $W=2$. In literature\cite{Lane2018}, the material is known to be a metal within DFT. When including Hubbard $U$, we observe a Mott insulating state. However, the planar $W=2$ altermagnetic spin-momentum locking  is present in both cases. 

Possible explanations why the spin-split altermagnetic phase  escaped attention are  that (i) previous studies focused on the high symmetry planes in the Brillouin zone, such as the grey-shaded $k_z=0$ plane in Fig.~S\ref{La2CuO4}b, whose spin degeneracy is protected by the spin group, or (ii) the spin splitting around the Fermi level on the $\sim 10$~meV scale is relatively weak. 
 
\begin{figure}[h!]
\begin{center}
\hspace*{-0cm}\epsfig{width=1\columnwidth,angle=0,file=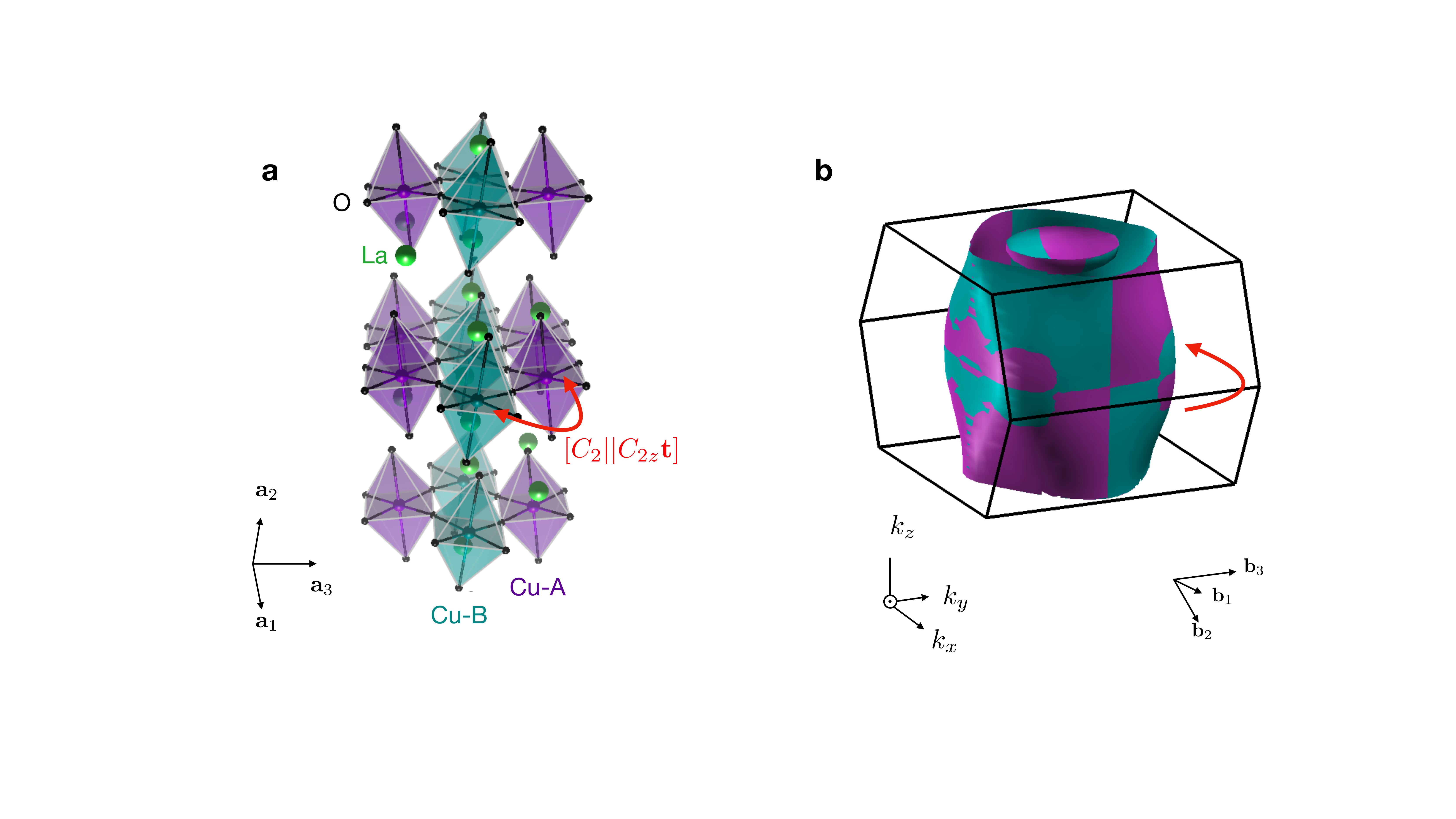}
\end{center}
\caption{{\bf{Altermagnetic parent cuprate  La$_2$CuO$_4$  of a high-temperature superconductor.}} {\bf a}, Schematic spin arrangement on the La$_2$CuO$_4$ crystal with opposite spin directions depicted by
purple and cyan color.  Red arrow and its label highlights the opposite-spin-sublattice transformation, containing a real-space two-fold
rotation. {\bf b}, Calculated  spin-momentum locking with the planar spin winding number  $W=2$  on top of the DFT  Fermi surface.  
}
\label{La2CuO4}
\end{figure}


\subsection*{Acknowledgements}
We acknowledge fruitful interactions with Igor Mazin, Rafael Gonz\'alez-Hern\'andez,  Helen Gomonay and Roser Valent\'{\i}. This work was supported by  Ministry of Education of the Czech Republic Grants LNSM-LNSpin, LM2018140,
Czech Science Foundation Grant No. 19-28375X,  EU FET Open RIA Grant No. 766566, SPIN+X (DFG SFB TRR 173) and Elasto-Q-Mat (DFG SFB TRR 288). We acknowledge the computing time granted on the supercomputer Mogon at Johannes Gutenberg University Mainz (hpc.uni-mainz.de).

\end{document}